\documentclass[twocolumn,english,aps,prl,superscriptaddress]{revtex4-1}
\usepackage[latin9]{inputenc}
\pdfoutput=1
\usepackage{amsmath}
\usepackage{amssymb}
\usepackage{graphicx}

\makeatletter

\providecommand{\tabularnewline}{\\}

 \@ifundefined{textcolor}{}
 {%
   \definecolor{BLACK}{gray}{0}
   \definecolor{WHITE}{gray}{1}
   \definecolor{RED}{rgb}{1,0,0}
   \definecolor{GREEN}{rgb}{0,1,0}
   \definecolor{BLUE}{rgb}{0,0,1}
   \definecolor{CYAN}{cmyk}{1,0,0,0}
   \definecolor{MAGENTA}{cmyk}{0,1,0,0}
   \definecolor{YELLOW}{cmyk}{0,0,1,0}
 }

\usepackage{babel}

\makeatother

\usepackage{babel}
\begin{document}

\title{Quantum many-body dynamics in optomechanical arrays}

\author{Max Ludwig}

\email{max.ludwig@physik.uni-erlangen.de}

\selectlanguage{english}%

\affiliation{Institute for Theoretical Physics, Universit\"at Erlangen-N\"urnberg,
Staudtstra\ss e 7, 91058 Erlangen, Germany}

\author{Florian Marquardt}

\affiliation{Institute for Theoretical Physics, Universit\"at Erlangen-N\"urnberg,
Staudtstra\ss e 7, 91058 Erlangen, Germany}

\affiliation{Max-Planck-Institute for the Science of Light, G\"unther-Scharowsky-Stra\ss e
1/Bau 24, 91058 Erlangen, Germany}
\begin{abstract}
We study the nonlinear driven dissipative quantum dynamics of an array
of optomechanical systems. At each site of such an array, a localized
mechanical mode interacts with a laser-driven cavity mode via radiation
pressure, and both photons and phonons can hop between neighboring
sites. The competition between coherent interaction and dissipation
gives rise to a rich phase diagram characterizing the optical and
mechanical many-body states. For weak intercellular coupling, the
mechanical motion at different sites is incoherent due to the influence
of quantum noise. When increasing the coupling strength, however,
we observe a transition towards a regime of phase-coherent mechanical
oscillations. We employ a Gutzwiller ansatz as well as semiclassical
Langevin equations on finite lattices, and we propose 
a realistic experimental implementation in optomechanical crystals.
\end{abstract}
\maketitle
\textit{Introduction. -} Recent experimental progress has brought
optomechanical systems into the quantum regime: A single mechanical
mode interacting with a laser-driven cavity field has been cooled
to the ground state \cite{2011_Teufel_SidebandCooling_Nature,2011_Chan_LaserCoolingNanomechOscillator}.
Several of these setups, in particular optomechanical crystals, offer
the potential to be scaled up to form optomechanical arrays. Applications
of such arrays for quantum information processing \cite{2011_Chang_SlowingAndStoppingLight_NJP,2012_Schmidt_OptomechanicalCircuits}
have been proposed. Given these developments, one is led to explore
quantum many-body effects in optomechanical arrays. In this work,
we analyze the nonlinear photon and phonon dynamics in a homogeneous
two-dimensional optomechanical array. In contrast to earlier works
\cite{2011_Chang_SlowingAndStoppingLight_NJP,2012_Schmidt_OptomechanicalCircuits,2012_Tomadin_ReservoirEngineering,2012_Xuereb_StrongCoupling},
here we study the array's quantum dynamics beyond a quadratic Hamiltonian.
To tackle the non-equilibrium many-body problem of this nonlinear
dissipative system, we employ a mean-field approach for the collective
dynamics. First, we discuss photon statistics in the array, in particular
how the photon blockade effect \cite{2011_Rabl_PhotonBlockade_PRL}
is altered in the presence of intercellular coupling. The main part
of the article focuses on the transition of the collective mechanical
motion from an incoherent state (due to quantum noise) to an ordered
state with phase-coherent mechanical oscillations. For these dynamics,
the dissipative effects induced by the optical modes play a crucial
role. On the one hand, they allow the mechanical modes to settle into
self-induced oscillations \cite{SelfOscillations}
once the optomechanical amplification rate exceeds the intrinsic mechanical
damping. On the other hand, the fundamental quantum noise (e.g. cavity
shot noise) diffuses the mechanical phases and prevents the mechanical
modes from synchronizing. This interplay leads to an elaborate phase
diagram characterizing the transition. We develop a semiclassical
model to describe the effective dynamics of the mechanical phases
and to study the system on finite lattices. 

While true long-range order is prohibited for a two-dimensional system
with continuous symmetry, at least for equilibrium systems, a Beresinskii-Kosterlitz-Thouless
transition towards a state with quasi-long-range order is possible.
The ordered mechanical phase thus resembles the superfluid phase in
two-dimensional cold atomic gases \cite{2006_Hadzibabic_BKTCrossover}
or Josephson junction arrays \cite{2001_Fazio_QPTAndVortexDynamics}.
Notably, optomechanical arrays combine the tunability of optical systems
with the robustness and durability of an integrated solid-state device.
Other driven dissipative systems that have been studied with regard
to phase transitions recently include cold atomic gases \cite{2008_Diehl,2010_Nagy_Dicke-ModelPhaseTransition,2010_Baumann_DickeQPT,2011_Lee_AntiferromagneticPhaseTransition},
nonlinear cavity arrays \cite{2006_Hartmann_StronglyInteractingPolaritons,2006_Greentree_QuPhaseTransitionsOfLight}
and optical fibers \cite{2008_Chang_CrystallizationOfStonglyInteractingPhotons}.
In a very recent work and along the lines of \cite{2008_Diehl}, the
preparation of long-range order for photonic modes was proposed using
the linear dissipative effects in an optomechanical array \cite{2012_Tomadin_ReservoirEngineering}.
Our work adds the novel aspect of a mechanical transition to the studies
of driven dissipative many-body systems.

\begin{figure}[t]
\includegraphics[width=0.95\columnwidth]{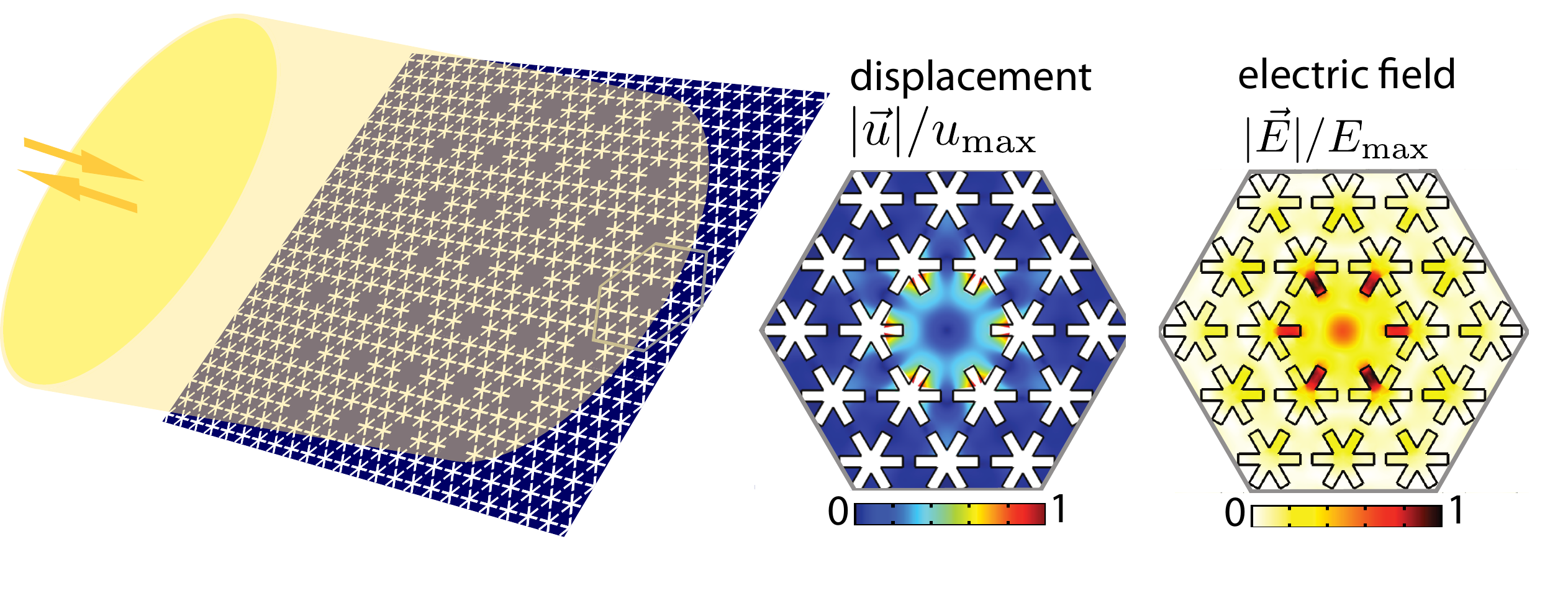}

\caption{Example implementation of an optomechanical array: A two-dimensional
snowflake optomechanical crystal \cite{2010_Safavi_2DSimultaneousBandgap,2011_Safavi-Naeini_PPT}
supports localized optical and mechanical modes around defect cavities.
Here, we propose arranging them in a super structure, forming the
array. The insets show electric field $\vec{E}$ and displacement
field $\vec{u}$ of an isolated defect cavity (obtained from finite
element simulations). Due to the finite overlap between modes of neighboring
sites \cite{2012_Ludwig_QuManyBody}, photons and phonons can hop
through the array, see Eq. (\ref{eq:H_int}). A wide laser beam drives
the optical modes of the array continuously and the reflected light
is read out.}

\label{Fig1} 
\end{figure}

\textit{Model. - }We study the collective quantum dynamics of a two-dimensional
homogeneous array of optomechanical cells (Fig. \ref{Fig1}). Each
of these cells consists of a mechanical mode and a laser driven optical
mode that interact via the radiation pressure coupling at a rate $g_{0}$
($\hbar=1$): 
\begin{equation}
\hat{H}_{{\rm om},j}=-\Delta\hat{a}_{j}^{\dagger}\hat{a}_{j}+\Omega\hat{b}_{j}^{\dagger}\hat{b}_{j}-g_{0}(\hat{b}_{j}^{\dagger}+\hat{b}_{j})\hat{a}_{j}^{\dagger}\hat{a}_{j}+\alpha_{L}(\hat{a}_{j}^{\dagger}+\hat{a}_{j}).\label{eq:H_single}
\end{equation}
The mechanical mode ($\hat{b}_{j}$) is characterized by a frequency
$\Omega.$ The cavity mode ($\hat{a}_{j}$) is transformed into the
frame rotating at the laser frequency ($\Delta=\omega_{{\rm laser}}-\omega_{{\rm cav}}$)
and driven at the rate $\alpha_{L}$. In the most general case, both
photons and phonons can tunnel between neighboring sites $\langle ij\rangle$
at rates $J/z$ and $K/z$, where $z$ denotes the coordination number.
The full Hamiltonian of the array is given by $\hat{H}=\sum_{j}\hat{H}_{{\rm om,}j}+\hat{H}_{{\rm int}}$,
with

\begin{eqnarray}
\hat{H}_{{\rm int}} & = & -\frac{J}{z}\sum_{\langle i,j\rangle}\big(\hat{a}_{i}^{\dagger}\hat{a}_{j}+\hat{a}_{i}\hat{a}_{j}^{\dagger}\big)-\frac{K}{z}\sum_{\langle i,j\rangle}\big(\hat{b}_{i}^{\dagger}\hat{b}_{j}+\hat{b}_{i}\hat{b}_{j}^{\dagger}\big).\label{eq:H_int}
\end{eqnarray}
To bring this many-body problem into a treatable form, we apply the
Gutzwiller ansatz $\hat{A}_{i}^{\dagger}\hat{A}_{j}\approx\langle\hat{A}_{i}^{\dagger}\rangle\hat{A}_{j}+\hat{A}_{i}^{\dagger}\langle\hat{A}_{j}\rangle-\langle\hat{A}_{i}^{\dagger}\rangle\langle\hat{A_{j}}\rangle$
to Eq. (\ref{eq:H_int}). The accuracy of this approximation improves
if the number of neighboring sites $z$ increases. For identical cells,
the index $j$ can be dropped and the Hamiltonian reduces to a sum
of independent contributions, each of which is described by

\begin{eqnarray}
\hat{H}_{{\rm mf}} & = & \hat{H}_{{\rm om}}-J\big(\hat{a}^{\dagger}\langle\hat{a}\rangle+\hat{a}\langle\hat{a}^{\dagger}\rangle\big)-K\big(\hat{b}^{\dagger}\langle\hat{b}\rangle+\hat{b}\langle\hat{b}^{\dagger}\rangle\big).\label{eq:H_mf}
\end{eqnarray}
Hence, a Lindblad master equation for the single cell density matrix
$\hat{\rho}$, $d\hat{\rho}/dt=-i[\hat{H}_{{\rm mf}},\hat{\rho}]+\kappa\mathcal{D}[\hat{a}]\hat{\rho}+\Gamma\mathcal{D}[\hat{b}]\hat{\rho}$
can be employed. The Lindblad terms $\mathcal{D}[\hat{A}]\hat{\rho}=\hat{A}\hat{\rho}\hat{A}^{\dagger}-\hat{A}^{\dagger}\hat{A}\hat{\rho}/2-\rho\hat{}\hat{A}^{\dagger}\hat{A}/2$
take into account photon decay at a rate $\kappa$ and mechanical
dissipation (here assumed due to a zero temperature bath) at a rate
$\Gamma$.

\textit{Photon statistics. - }Recently, it was shown that the effect
of photon blockade \cite{2011_Rabl_PhotonBlockade_PRL} can appear
in a single optomechanical cell: The interaction with the mechanical
mode induces an optical nonlinearity of strength $g_{0}^{2}/\Omega$
\cite{2011_Rabl_PhotonBlockade_PRL,2011_Nunnenkamp_SinglePhotonOptomechanics_PRL}
and the presence of a single photon can hinder other photons from
entering the cavity. To observe this effect, the nonlinearity must
be comparable to the cavity decay rate, i.e. $g_{0}^{2}/\Omega\gtrsim\kappa$,
and the laser drive weak ($\alpha_{L}\ll\kappa$) \cite{2011_Rabl_PhotonBlockade_PRL,2013_Kronwald_FullPhotonStatistics}.
\begin{figure}[t]
\includegraphics[width=0.95\columnwidth]{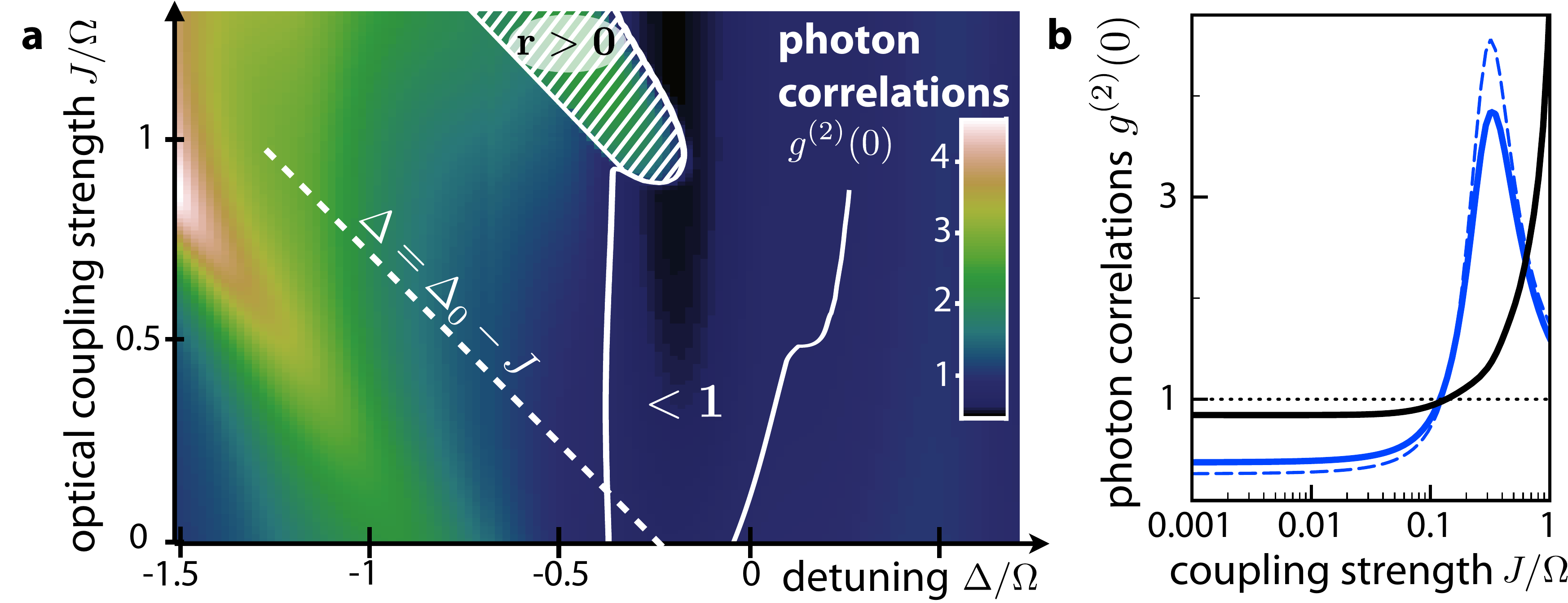}

\caption{Loss of photon blockade for increasing optical coupling in an array
of optomechanical cavities. (a) The equal time photon correlation function
shows anti-bunching $(g^{(2)}(0)<1)$ and bunching $(g^{(2)}(0)>1)$
as a function of detuning $\Delta$ and optical coupling strength
$J$. The smallest values of $g^{(2)}(0)$ are found for a detuning
$\Delta_{0}=-g_{0}^{2}/\Omega$. (b) When increasing the coupling $J$
while keeping the intracavity photon number constant, i.e. along the
dashed line in panel (a), photon blockade is lost (black solid line). 
For a smaller driving power (blue solid line, $\alpha_{L}=5\cdot10^{-5}\kappa$), 
anti-bunching is more pronounced and the behavior is comparable to that of 
a nonlinear cavity (dashed line). The hatched area in (a) outlines
a region where a transition towards coherent mechanical oscillations
has set in. $\kappa=0.3\,\Omega$, $\alpha_{L}=0.65\,\kappa$, $g_{0}=0.5\,\Omega$,
$\Gamma=0.074\,\Omega$. }

\label{Fig2} 
\end{figure}

To study nonclassical effects in the photon statistics, we analyze
the steady-state photon correlation function $g^{(2)}(\tau)=\langle\hat{a}^{\dagger}(t)\hat{a}^{\dagger}(t+\tau)\hat{a}(t+\tau)\hat{a}(t)\rangle/\langle\hat{a}(t)^{\dagger}\hat{a}(t)\rangle^{2}$
\cite{1995_MandelWolf_OpticalCoherence_Book} at equal times ($\tau=0$).
Here (Fig. \ref{Fig2}), we probe the influence of the collective
dynamics by varying the optical coupling strength $J$, while keeping
the mechanical coupling $K$ zero for clarity. We note that, when
increasing $J$, the optical resonance effectively shifts: $\Delta\to\Delta+J$.
To keep the photon number fixed while increasing $J$, the detuning
has to be adapted \cite{2012_Nissen_Qubit-CavityArrays}. In this
setting, we observe that the interaction between the cells suppresses
anti-bunching (Fig. \ref{Fig2} (b)). Photon blockade is lost
if the intercellular coupling becomes larger than the effective nonlinearity,
$2J\gtrsim g_{0}^{2}/\Omega$. Above this value, the photon statistics
shows bunching, and ultimately reaches Poissonian statistics for large
couplings. Similar physics has recently been analyzed for coupled
qubit-cavity arrays, \cite{2012_Nissen_Qubit-CavityArrays}. For large
coupling strengths, though, Fig. \ref{Fig2}(a) reveals
signs of the collective mechanical motion (hatched area). There we
observe the correlation function to oscillate (at the mechanical frequency)
and to show bunching. We will now investigate this effect.

\textit{Collective mechanical quantum effects. -} To describe the
collective mechanical motion of the array, we focus on the case of
purely mechanical intercellular coupling ($K>0$, $J=0$) for simplicity.
Note, though, that the effect is also observable for optically coupled
arrays, as discussed above.

\begin{figure}
\includegraphics[width=0.95\columnwidth]{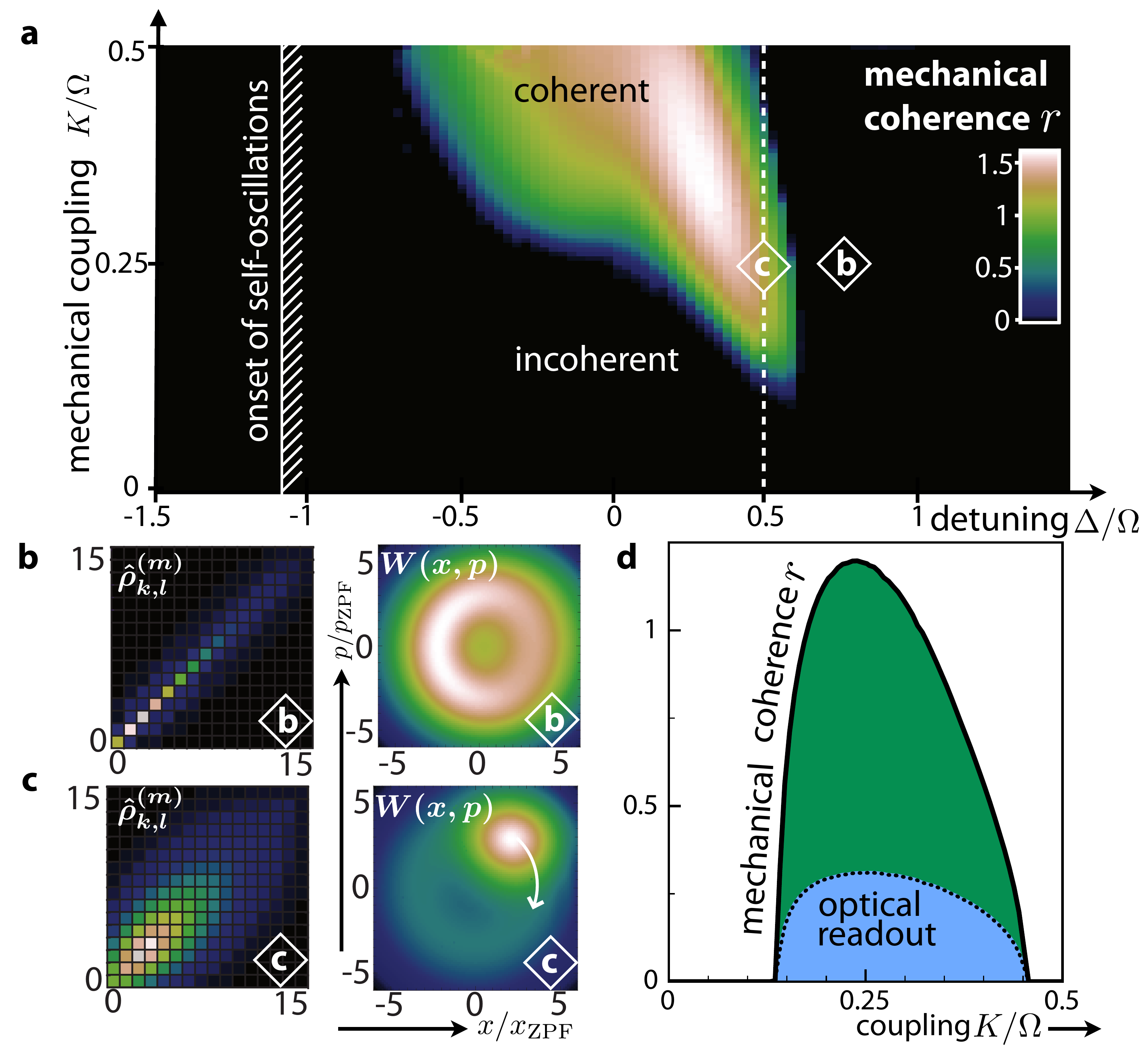}

\caption{Transition from the incoherent to the synchronized (coherent) phase:
(a) Mechanical coherence $r$ (Eq. (\ref{eq:orderparam})) as a function
of laser detuning $\Delta$ and mechanical coupling $K$. At weak
coupling, the self-oscillations are incoherent, $r=0$, due to quantum
noise. When increasing the coupling strength, the systems shows a
sharp transition towards the ordered regime, where the mechanical
oscillations are phase-coherent, $r>0$. (b,c) Modulus of the density
matrix elements (in Fock space) and Wigner density of the collective
mechanical state in the incoherent (b) and the coherent regime (c),
as marked in (a). (d) Mechanical coherence $r$ as a function of
coupling strength $K$ along the dashed line in (a). The dotted line
shows the optical readout of coherence, i.e. the oscillating component
of the photon number $\langle\hat{a}^{\dagger}\hat{a}\rangle$, proportional
to the intensity of the reflected beam and thus directly accessible
in experiment. $g_{0}=\kappa=0.3\,\Omega$, $\alpha_{L}=1.1\,\kappa$,
$\Gamma=0.074\,\Omega$}

\label{Fig3} 
\end{figure}

As our main result, Figs. \ref{Fig3}(a) and (d) show the sharp transition
between incoherent self-oscillations and a phase-coherent collective
mechanical state as a function of both laser detuning $\Delta$ and
coupling strength $K$: In the regime of self-induced oscillations,
the phonon number $\langle\hat{b}^{\dagger}\hat{b}\rangle$ reaches
a finite value. Yet, the expectation value $\langle\hat{b}\rangle$
remains small and constant in time. When increasing the intercellular
coupling, though, $\langle\hat{b}\rangle$ suddenly starts oscillating
and reaches a steady state

\begin{equation}
\langle\hat{b}\rangle(t)=\bar{b}+re^{-i\Omega_{{\rm eff}}t}.\label{eq:orderparam}
\end{equation}
Here, we introduced the mechanical coherence $r$ and the oscillation
frequency $\Omega_{{\rm eff}}$, which is shifted by the optical fields
and the intercellular coupling, cf. Eq. (\ref{eq:PhaseEq}).

Our more detailed analysis (see below) indicates that this transition
results from the competition between the fundamental quantum noise
of the system and the tendency of phase locking between the coupled
nonlinear oscillators. Below threshold, the quantum noise from the
phonon bath and the optical fields diffuses the mechanical phases
at different sites and drives the mechanical motion into an incoherent
mixed state. The reduced density matrix $\hat{\rho}^{(m)}$ is predominantly
occupied on the diagonal, see Fig. \ref{Fig3}(b), and the Wigner
distribution, $W(x,p)=\frac{1}{\pi\hbar}\int_{-\infty}^{\infty}\langle x-y|\hat{\rho}^{(m)}|x+y\rangle e^{2ipy/\hbar}dy$,
has a ringlike shape, reflecting the fact that the mechanical phase
is undetermined \cite{2008_ML_OptomechInstab,2012_Qian_QuSignatures}.
Above threshold, the mechanical motion at different sites becomes
phase locked, and the coherence parameter $r$ reaches a finite value.
The emergence of coherence also becomes apparent from the off-diagonal
elements of $\hat{\rho}^{(m)}$ (Fig. \ref{Fig3}(c)). The corresponding
Wigner function assumes the shape of a coherent state with a definite
phase oscillating in phase space. Thus, this transition spontaneously
breaks the time-translation symmetry. In a two-dimensional implementation,
true long-range order is excluded, but the coherence between different
sites is expected to decay as a power law with distance. We also note
that this transition is the quantum mechanical analogon of classical
synchronization, which was studied for optomechanical systems in \cite{2011_Heinrich_CollectiveDynamics,2012_Holmes_Synchronization,2012_Zhang_SynchronizationOptomech}.
An important difference is, though, that the classical nonlinear dynamics
was analyzed for an inhomogeneous (with disordered mechanical frequencies)
system in the absence of noise \cite{2011_Heinrich_CollectiveDynamics,2012_Holmes_Synchronization,2012_Zhang_SynchronizationOptomech},
while in our case disorder is only introduced via fundamental quantum
noise. Quantum synchronization has also been discussed in the context
of linear oscillators \cite{2012_Giorgi_QuCorrelations} and nonlinear
cavities \cite{2012_Lee_QuSynchronization} recently.

The laser detuning determines both the strength of the self-oscillations
and the influence of the cavity shot noise on the mechanical motion.
It turns out that the diffusion of the mechanical phases is pronounced
close to the onset of self-oscillations and at the mechanical sideband
\cite{2012_Ludwig_QuManyBody}. As we will show below, even the coherent
coupling between the mechanical phases (ultimately leading to synchronization)
is tunable via the laser frequency. As a result, the synchronization
threshold depends non-trivially on the detuning parameter $\Delta$,
see Fig. \ref{Fig3}(a).

\textit{Langevin dynamics on finite lattices. -} In order to gain further
insight into the coupling and decoherence mechanisms as well as effects
of geometry and dimensionality, we analyze the semi-classical Langevin
equations of the full optomechanical array:
\begin{eqnarray}
\dot{\beta}_{i} & = & \big(-i\Omega-\frac{\Gamma}{2}\big)\beta_{i}+ig_{0}|\alpha_{i}|^{2}+i\frac{K}{z}\sum_{\langle ij\rangle}\beta_{j}+\sqrt{\frac{\Gamma}{2}}\xi_{\beta}\nonumber \\
\dot{\alpha}_{i} & = & \Big(i\Delta+ig_{0}(\beta_{i}+\beta_{i}^{*})-\frac{\kappa}{2}\Big)\alpha_{i}-i\alpha_{L}+\sqrt{\frac{\kappa}{2}}\xi_{\alpha}.\label{eq:EoM}
\end{eqnarray}

The fluctuating noise forces $\xi_{\sigma=\alpha,\beta}(t)$ mimic
the effects of the zero temperature phonon bath and the cavity shot
noise, respectively. They are independent at each site and obey $\langle\xi_{\sigma}\rangle=0$
and $\langle\xi_{\sigma}(t)\xi_{\sigma}^{*}(t')\rangle=\delta(t-t')$.
In this context, $\langle...\rangle$ denotes the average over
different realizations of the stochastic terms.
This Langevin approach is equivalent to the truncated
Wigner approximation (see  \cite{2010_Polkovnikov_ReviewTWA} for a review), 
and it has shown good qualitative agreement with the full quantum
dynamics for a single optomechanical cell \cite{2008_ML_OptomechInstab,2010_Rodrigues_AmplitudeNoiseSuppression}. It allows us
to treat the effects of quantum fluctuations at all wavelengths on the
spatial phase correlations via numerical simulations. At this point, a full
quantum treatment for sufficiently large systems remains
a challenging problem for future studies.

\begin{figure}[t]
\includegraphics[width=0.95\columnwidth]{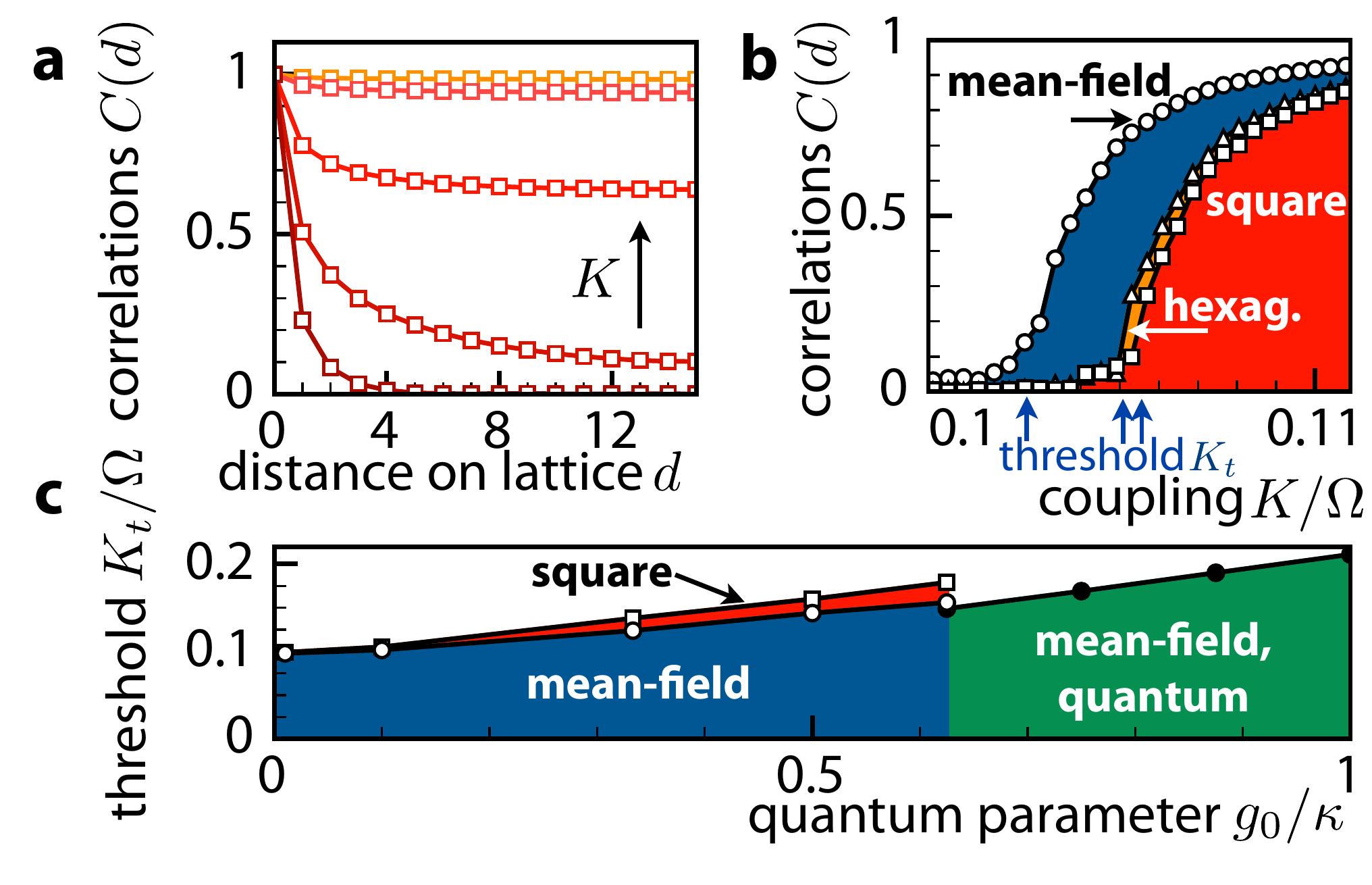}\caption{Langevin dynamics on finite lattices: (a) Correlations
$C(d=|i-j|)=|\langle e^{i\varphi_{i}}e^{-i\varphi_{j}}\rangle|$ in
a $30\times30$ optomechanical array. Quasi-long-range order sets
in for sufficiently large coupling strengths. $K=\{0.09,0.105,0.107,0.12,0.15\}\Omega$
(b) Correlations over a distance of $d=14$ as a function of mechanical
coupling strength $K$ for a square lattice ($z=4,$ squares), a hexagonal
lattice ($z=6,$ triangles) slightly below the mean-field result (circles).
(c) Coupling threshold as a function of quantum parameter $g_{0}/\kappa$
(squares: square lattice, empty (filled) circles: semi-classical (quantum)
mean-field approach). $\Delta+g_{0}^{2}/\Omega=0.34$, $g_{0}=0.1\kappa$
in (a),(b), $g_{0}\alpha_{L}={\rm 0.33\,\kappa}$ in (c), other parameters
as in Fig. \ref{Fig3}.}

\label{Fig4} 
\end{figure}

First, we study the onset of quasi-long-range order in a finite system.
To this end we evaluate the correlations $C(d=|i-j|)=\langle e^{i\varphi_{i}}e^{-i\varphi_{j}}\rangle$,
where $e^{i\varphi_{i}}=\beta_{i}/|\beta_{i}|$. Numerical calculations
on a $30\times30$ square lattice (see Fig. \ref{Fig4}(a)) indicate
that for weak intercellular coupling the mechanical phases at different
sites are uncorrelated even for small distances $d$. When increasing
the coupling strength, however, the mechanical motion becomes correlated
over the whole array with only a slow decrease with distance. The
coupling threshold, here defined by setting a lower bound of $C(14)>0.01$,
varies with coordination number, see Fig. \ref{Fig4}(b). Within the
mean-field approximation, i.e. for a lattice with global coupling
of all sites, fluctuations between neighboring sites and hence the
threshold value are underestimated. The coupling threshold grows with
the quantum parameter \cite{2008_ML_OptomechInstab}, i.e. the ratio
of optomechanical coupling and cavity decay rate, $g_{0}/\kappa$,
see Fig. \ref{Fig4}(c): For $g_{0}\approx\kappa$, single photons
and phonons interact strongly and quantum fluctuations hamper synchronization.

\textit{Synchronization threshold. - }For an analytical approach,
the complexity of the Langevin equations can be reduced by integrating
out the dynamics of the optical modes and the mechanical amplitudes
and by going back to the mean-field approximation \cite{2012_Ludwig_QuManyBody}.
The resulting equation describes the coupling of the mechanical phase
on a single site, $\varphi$, to a mean field $\mbox{\ensuremath{\Psi}}$:

\begin{eqnarray}
\dot{\varphi} & = & -\Omega(\bar{A})+KR\cos(\Psi-\varphi)+K_{1}R\sin(\Psi-\varphi)\nonumber \\
 &  & +\sqrt{2D_{\varphi}}\xi_{\varphi}+\mbox{\ensuremath{\mathcal{O}}}(R^{2}).\label{eq:PhaseEq}
\end{eqnarray}
Here, the order parameter is defined as $\langle e^{i\varphi_{j}}\rangle\equiv Re^{i\Psi}$.
The rate $K_{1}=(d\Omega-K/2)K/\gamma$ determines the coupling of
phases mediated by slow amplitude modulations between neighboring
sites. These beat modes couple back to the phase dynamics via the
amplitude dependent optical spring effect, $\Omega(\mbox{\ensuremath{\bar{A}}})+d\Omega\cdot(A-\bar{A})/\bar{A},$
where $d\Omega=\bar{A}\frac{d\Omega}{dA}|_{A=\bar{A}}$, and the bare
mechanical coupling $K$, leading to two opposing terms in $K_{1}$.
Here, $\bar{A}$ denotes the steady state mechanical amplitude and
$\gamma$ the amplitude decay rate set by the optical field. The fluctuating
noise force $\sim\xi_{\varphi}$ comprises the effects of mechanical
fluctuations and radiation pressure noise and is characterized by
a diffusion constant $D_{\varphi}$ \cite{2010_Rodrigues_AmplitudeNoiseSuppression,2012_Ludwig_QuManyBody}.

Equation (\ref{eq:PhaseEq}) reveals the close connection to the Kuramoto
model \cite{1984_Kuramoto_CooperativeDynamics} and the two-dimensional
xy-model. In the incoherent regime, the order parameter $R$ is zero
and the phase fluctuates freely. In the coherent regime, the restoring
force $\sim K_{1}R$ leads the phase $\varphi$ towards a fixed relation
with $\Psi$. The cosine term only renormalizes the oscillation frequency.
This statement can be clarified by a linear stability analysis, see
\cite{1991_Strogatz_StabilityOfIncoherence,2012_Ludwig_QuManyBody}.
It turns out that the incoherent phase becomes unstable for 
\begin{equation}
K_{1}=2D_{\varphi},\label{eq:threshold}
\end{equation}
defining the threshold of the transition. Moreover, if $K_{1}$ becomes
negative, no stable phase synchronization is possible. This situation
arises if $d\Omega<0$, or for large intercellular coupling rates
$K>2d\Omega$, see Fig. \ref{Fig3}(d).

\textit{Experimental prospects. - }We note that observation of the
mechanical phase transition does \emph{not} require single photon
strong coupling ($g_{0}\gtrsim\kappa$): The quantum fluctuations
of the light field will dominate over thermal fluctuations as long
as $4g_{0}^{2}|\alpha|^{2}/\kappa>k_{B}T/Q$. This is essentially
the condition for ground-state cooling, which has been achieved using
high-$Q$ mechanical resonators and cryogenic cooling \cite{2011_Teufel_SidebandCooling_Nature,2011_Chan_LaserCoolingNanomechOscillator},
see Table \ref{Table1}. In contrast, the photon-blockade effect (Fig.
\ref{Fig2}) requires low temperatures $T$ and $g_{0}^{2}\gtrsim\Omega\kappa$,
or at least, in a slightly modified setup \cite{2012_Stannigel_OptomechanicalQIP,2012_ML_EnhancedQuNonlinearities},
$g_{0}\gtrsim\kappa$. While still challenging, optomechanical systems
are approaching this regime \cite{2012_Chan_OptimizedOptomechanicalCrystalCavity}.
\begin{table}[b]
\begin{tabular}{|c|c|c|c|c|c|}
\hline 
Setup & $T\:[K]$  & $\Gamma n_{{\rm th}}/\Omega$ & $\Gamma_{{\rm opt}}/\Omega$ & $g_{0}/\kappa$ & $L\:[\mu m]$\tabularnewline
\hline 
\hline 
Microwave based  & $25\,{\rm mK}$ & $10^{-4}$ & $7\times10^{-3}$ & $10^{-3}$ & $\sim100$\tabularnewline
\hline 
Optomech. crystal & $20\,{\rm K}$ & $10^{-3}$ & $4\times 10^{-3}$ & $2\cdot10^{-3}$ & $\sim4$\tabularnewline
\hline 
Microtoroid & $650\,{\rm mK}$ & $7\times10^{-2}$ & $6\times 10^{-2}$ & $5\times10^{-4}$ & $\sim30$\tabularnewline
\hline 
\end{tabular}

\caption{Parameters of optomechanical systems \cite{2011_Teufel_SidebandCooling_Nature,2011_Chan_LaserCoolingNanomechOscillator,2012_Verhagen_QuantumCoherentCoupling}:
Temperature of phonon bath $T$, strength of mechanical fluctuations
$\Gamma n_{{\rm th}}\approx k_{B}T/Q$, strength of cavity shot noise
$\Gamma_{{\rm opt}}\approx4g_{0}^{2}|\bar{\alpha}|^{2}/\kappa$, quantum
parameter $g_{0}/\kappa$ and approximate size $L$.}

\label{Table1}
\end{table}

Microfabricated optomechanical systems such as microresonators (e.g.
\cite{2012_Verhagen_QuantumCoherentCoupling}), optomechanical crystals
(e.g. \cite{2011_Chan_LaserCoolingNanomechOscillator}) or microwave-based
setups (e.g. \cite{2011_Teufel_SidebandCooling_Nature}) lend themselves
to extensions to optomechanical arrays. Here, we focus on optomechanical
crystals, which are well suited due to their extremely small mode
volumes. The properties of two-dimensional optomechanical crystals
have been analyzed in \cite{2010_Safavi_2DSimultaneousBandgap}. The
finite overlap of the evanescent tails of adjacent localized modes
\cite{2012_Ludwig_QuManyBody} results in a coupling of the form of
Eq. (\ref{eq:H_int}), in analogy to the tight-binding description
of electronic states in solids. Sufficiently strong optical and mechanical
hopping rates are feasible, see \cite{2011_Heinrich_CollectiveDynamics}
for one-dimensional and \cite{2012_Ludwig_QuManyBody} for two-dimensional
structures. The simultaneous optical driving of many cells may be
realized by a single broad laser beam irradiating the slab, see Fig.
\ref{Fig1}. Alternatively, similar physics may be observed for many
mechanical modes coupling to one extended in-plane optical mode \cite{2011_Heinrich_CollectiveDynamics,2012_Holmes_Synchronization,2012_Xuereb_StrongCoupling}
(thereby effectively realizing global coupling). 

The transition towards the synchronized phase can be detected by probing
the light reflected from the optomechanical array and measuring the
component oscillating at the mechanical frequency, see Fig. \ref{Fig3}(d).
To read out correlations between individual sites, the intensities
of individual defect cavities may be analyzed \cite{2012_Ludwig_QuManyBody},
for example by evanescently coupling them to tapered fibers or waveguides.

We expect the transition to be robust against disorder \cite{2011_Heinrich_CollectiveDynamics}.
One may also study the formation of vortices and other topological
defects induced by engineered irregularities and periodic variations,
and explore various different lattice structures or the possibility
of other order phases (e.g. anti-ferromagnetic order). Thus, optomechanical
arrays provide a novel, integrated and tunable platform for studies
of quantum many body effects.

The authors would like to thank Oskar Painter and Bj\"orn Kubala for
valuable discussions. This work was supported by the DFG Emmy-Noether
program, an ERC starting grant, the DARPA/MTO ORCHID program and the
ITN network cQOM.

\clearpage

\global\long\def\theequation{S.\arabic{equation}}
 \global\long\def\thefigure{S\arabic{figure}}

\thispagestyle{empty}
\onecolumngrid
\begin{center}
{\fontsize{12}{12}\selectfont \textbf{Supplemental Material for "Quantum many-body dynamics
in optomechanical arrays"\\[5mm]}}
{\normalsize Max Ludwig$^{1,*}$ and Florian Marquardt$^{1,2}$\\[1mm]}
{\fontsize{9}{9}\selectfont
\textit{	$^1$Institute for Theoretical Physics, Universit\"at Erlangen-N\"urnberg,
Staudtstra\ss e 7, 91058 Erlangen, Germany\\
 			$^2$Max-Planck-Institute for the Science of Light, G\"unther-Scharowsky-Stra\ss e
1/Bau 24, 91058 Erlangen, Germany}}
\vspace*{6mm}
\end{center}
\normalsize
\twocolumngrid

\section{\textup{Mean-field phase equation}}

In this section, we provide details for the derivation of the mean-field
phase equation, Eq. (\ref{eq:PhaseEq}), starting from the equations
of motion of the optomechanical array, Eqs. (\ref{eq:EoM}). 
Introducing phases $\varphi_{j}$ and amplitudes $A_{j}$ (in units
of the mechanical ground state width) as new coordinates of mechanical
motion and omitting fast oscillating terms, the so-called Hopf equations
are derived directly from (\ref{eq:EoM}). We follow \cite{2011_Heinrich_CollectiveDynamics},
but add noise terms:

\begin{eqnarray}
\dot{\varphi_{i}} & = & -\Omega(A)+\frac{K}{zA_{i}}\sum_{\langle ij\rangle}A_{j}\cos(\varphi_{j}-\varphi_{i})+\frac{\tilde{\xi}_{\varphi}}{A_{i}}\nonumber \\
\dot{A}_{i} & = & -\gamma\,(A_{i}-\bar{A})-\frac{K}{z}\sum_{\langle ij\rangle}A_{j}\sin(\varphi_{j}-\varphi_{i})+\tilde{\xi}_{A}.\label{eq:Hopf}
\end{eqnarray}
The steady amplitude $\bar{A}$ and the amplitude decay rate $\gamma$
are determined by the optical field: $\gamma(A-\bar{A})=\big(\Gamma+\Gamma_{{\rm opt}}(A)\big)A/2,$
where $\Gamma_{{\rm opt}}=-4g_{0}\langle|\alpha|^{2}\sin\varphi\rangle_{T}$ and
where $\langle...\rangle_{T}$ denotes the average over one mechanical
period. The mechanical oscillation frequency is modified via an amplitude
dependent optical spring effect: $\Omega(A)=\Omega-2g_{0}\langle|\alpha|^{2}\cos\varphi\rangle_{T}/A$.
The fluctuating noise forces $\tilde{\xi}_{\varphi}$ and $\tilde{\xi}_{A}$
comprise the effects of the phonon bath and the cavity shot noise,
see \cite{2010_Rodrigues_AmplitudeNoiseSuppression} and below. 

For weak coupling, $K/z\ll\Omega$, the fluctuations of the mechanical
amplitudes around the steady state value $\bar{A}$ are given by \cite{2011_Heinrich_CollectiveDynamics}
\begin{equation}
\delta A_{i}(t)\approx-\frac{K\bar{A}}{z\gamma}\sum_{\langle ij\rangle}\sin(\varphi_{j}(t)-\varphi_{i}(t)).\label{eq:deltaA(t)_gen}
\end{equation}
These beat modes introduce an effective second order coupling between
phases at different sites, as can be seen after plugging Eq. (\ref{eq:deltaA(t)_gen})
into the Hopf equation for $\varphi_{i}$ (\ref{eq:Hopf}) and performing
the time averages $\langle...\rangle_{T}$: 
\begin{eqnarray}
\dot{\varphi}_{i} & = & -\Omega(\bar{A})+\frac{K}{z}\sum_{\langle ij\rangle}\cos(\varphi_{j}-\varphi_{i})+\frac{K\, d\Omega}{z\,\gamma}\sum_{\langle ij\rangle}\sin(\varphi_{j}-\varphi_{i})\nonumber \\
 &  & +\frac{K^{2}}{2z^{2}\gamma}\sum_{\langle ij\rangle}\sum_{\langle jk\rangle}\big(\sin(2\varphi_{j}-\varphi_{k}-\varphi_{i})-\sin(\varphi_{k}-\varphi_{i})\big)\nonumber \\
 &  & +\frac{K^{2}}{2z^{2}\gamma}\sum_{\langle ij\rangle}\sum_{\langle ik\rangle}\sin(\varphi_{j}+\varphi_{k}-2\varphi_{i})+\xi_{\varphi},\label{eq:Kura0}
\end{eqnarray}
where we introduced $d\Omega=\bar{A}\frac{d\Omega}{dA}|_{A=\bar{A}}$.
This equation is similar to the xy model and the Kuramoto model in the 
presence of noise, but with additional terms that mainly shift the frequency (the $\cos$-term)
and indicate higher-order coupling (the contributions of the double
sums).

Ultimately, we apply a mean-field approximation: We replace $e^{i\varphi_{j}}$
for neighboring cells by $\langle e^{i\varphi_{j}}\rangle\equiv Re^{i\Psi}$
and $e^{i2\varphi_{j}}$ by $\langle e^{i2\varphi_{j}}\rangle\equiv R_{2}e^{i\Psi_{2}}$,
where $\langle...\rangle$ denotes the average over all sites \cite{1984_Kuramoto_CooperativeDynamics},
and arrive at the effective phase equation, Eq. (\ref{eq:PhaseEq}),
with additional second order contributions:

\begin{multline}
\dot{\varphi}=-\Omega(\bar{A})+KR\cos(\Psi-\varphi)+K_1R\sin(\Psi-\varphi)\\
+K_{2}R^{2}\sin(2\Psi-2\varphi)+K_{2}RR_{2}\sin(\Psi_{2}-\Psi-\varphi)+\xi_{\varphi},\label{eq:PhaseEq}
\end{multline}
where we introduced the effective coupling rates
\begin{eqnarray}
K_1=Kd\Omega/\gamma-K_{2},\\
K_2=K^2/2z^2\gamma.\label{eq:K12}
\end{eqnarray}

\section{Phase Diffusion}

Here, we list some more details of the phase diffusion in the system
based on the analysis given by Rodrigues and Armour \cite{2010_Rodrigues_AmplitudeNoiseSuppression}
for a single optomechanical cell. The diffusion constant associated
with $\xi_{\varphi}$ (see Eq. (\ref{eq:Kura0})) is given by 
\begin{equation}
D_{\varphi}=\frac{1}{\bar{A}^{2}}\big(\tilde{D}_{\varphi}+\frac{\delta\Omega^{2}}{\gamma^{2}}\tilde{D}_{A}\big),\label{eq:Diff_full}
\end{equation}
where $\tilde{D}_{\varphi}$ and $\tilde{D}_{A}$ correspond to the
noise acting on phase and amplitude in Eqs. (\ref{eq:Hopf}), and
where the diffusion rates are defined as 
\begin{equation}
2\tilde{D}_{\varphi,A}=\int_{-\infty}^{\infty} d\tau\langle\tilde{\xi}_{\varphi,A}(t+\tau)\tilde{\xi}_{\varphi,A}(t)\rangle_{T}\label{eq:Diff_def}
\end{equation}
and likewise for $D_{\varphi}$. For the explicit expressions we refer
to \cite{2010_Rodrigues_AmplitudeNoiseSuppression}. In the limit
of $g_{0}\bar{A}\ll\Omega$, and for $\Omega\gg\kappa$ and $\Delta=\Omega$,
the maximum diffusion rates can be approximated by 
\begin{equation}
2\tilde{D}_{\varphi,A}\approx\Gamma+\Gamma_{{\rm opt}},\label{eq:Diff_approx}
\end{equation}
i.e. the sum of the intrinsic mechanical damping $\Gamma$ and the
optomechanical damping rate at the mechanical sideband 
\begin{equation}
\Gamma_{{\rm opt}}\approx4g_{0}^{2}|\bar{\alpha}|^{2}/\kappa.\label{eq:Gamma_opt}
\end{equation}

\begin{figure}[t]
\includegraphics[width=1\columnwidth]{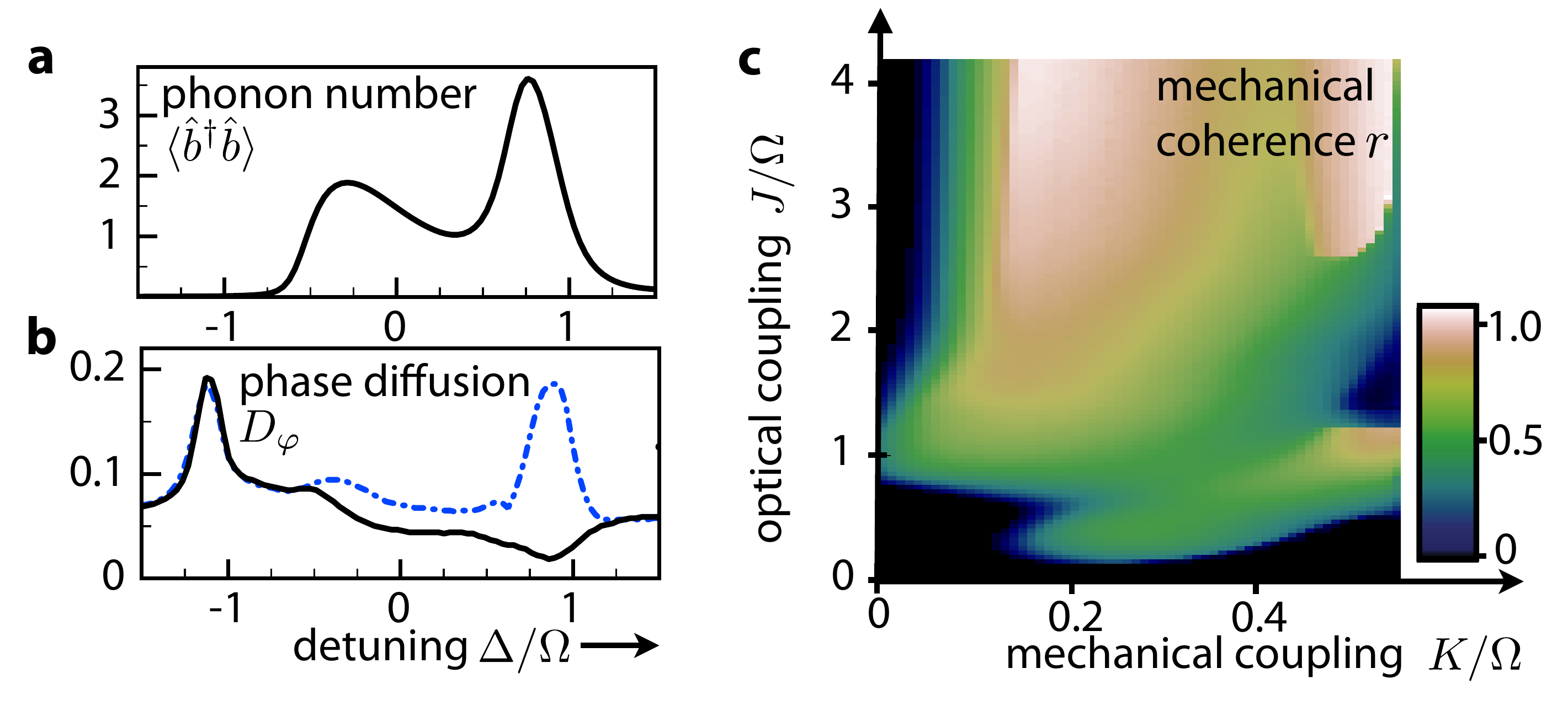}\caption{Additional details for the quantum dynamics of the optomechanical
array within the mean-field approximation: (a) The phonon number $\langle\hat{b}^{\dagger}\hat{b}\rangle$
shows, as a function of detuning, maxima at the resonance and at the
sideband ($\Delta\approx\Omega-g_{0}^{2}/\Omega$). (b) The diffusion
constant for the mechanical phase, $D_{\varphi}$, for an uncoupled
($K=0$, solid line) and a coupled array ($K=0.1\,\Omega$, dash-dotted
line). Other parameters as in Fig. \ref{Fig3}. (c) Interplay of mechanical
and optical coupling: Mechanical coherence is observed as a function
of both mechanical and optical coupling. $\Delta+J=\Omega/2$, other parameters
as in Fig.~\ref{Fig2}.}

\label{FigS1} 
\end{figure}

The diffusion of the mechanical phase can also be studied using the
full quantum simulations and evaluating the linewidth of the correlator
$\langle\hat{b}(t)\hat{b}^{\dagger}(0)\rangle\sim e^{-(i\Omega_{{\rm eff}}+D_{\varphi})t}$,
see Fig. \ref{FigS1}(b). Close to the onset, for small amplitudes
$\bar{A}$ and weak amplitude damping $\gamma$, the mechanical phases
are very susceptible to quantum noise preventing synchronization.
For finite coupling strengths, the diffusion is also enhanced, most
strikingly at the mechanical sideband. As a result, the synchronization
threshold shows a minimum between the onset of self-oscillations and
the sideband, as observed in Fig. \ref{Fig3}(a). From extended simulations,
we find that this behavior is generic for systems in the resolved
sideband regime ($\Omega>\kappa$).

\section{Stability analysis}

Here, we briefly recall details of the stability analysis leading
to Eq. (\ref{eq:threshold}), which, for the case of the Kuramoto model,
has been given in \cite{1988_Sakaguchi,1991_Strogatz_StabilityOfIncoherence}.
We consider the density of the mechanical phases, $\varrho(\varphi)$.
It is normalized, $\int_{0}^{2\pi}\varrho(\varphi)=1$, and the order
parameter $R$ and the mean-field $\Psi$ can be computed from $Re^{i\Psi}=\int_{0}^{2\pi}e^{i\varphi}\varrho(\varphi)d\varphi$
(and likewise for $R_{2}$ and $\Psi_{2}$). The Fokker-Planck equation
corresponding to Eq. (\ref{eq:PhaseEq}) is given by 
\begin{equation}
\partial_{t}\varrho+\partial_{\varphi}\big(\varrho v\big)=D_{\varphi}\partial_{\varphi}^{2}\varrho\label{eq:Fokker-Planck}
\end{equation}
with a velocity 
\begin{equation}
v=-\Omega(\bar{A})+K\cos(\Psi-\varphi)+K_{1}R\sin(\Psi-\varphi)+\mathcal{O}(R^{2}),\label{eq:FP_velocity}
\end{equation}
where second order contributions can be neglected for this linear
analysis. In the unsynchronized regime, the mechanical phases are
equally distributed over the interval $[0,2\pi]$, and $\varrho=(2\pi)^{-1}$.
To study the time evolution of a small fluctuation on top of the incoherent
background, we employ the ansatz \cite{1991_Strogatz_StabilityOfIncoherence}:
\begin{equation}
\varrho(\varphi)=\frac{1}{2\pi}+c(t)e^{i\varphi}+c^{*}(t)e^{-i\varphi},\label{eq:rho_ans}
\end{equation}
leading to
\begin{equation}
\dot{c}=-\Big(i\big(\Omega-\frac{K}{2}\big)+\big(D_{\varphi}-\frac{K_{1}}{2}\big)\Big)c.\label{eq:cdot}
\end{equation}
This equation reveals that the incoherent solution $\varrho=(2\pi)^{-1}$
becomes unstable for 
\begin{equation}
K_{1}\mbox{=}2D_{\varphi}.\label{eq:threshold-1}
\end{equation}
and thus defines the coupling threshold for the synchronization transition
\cite{1991_Strogatz_StabilityOfIncoherence}.

\section{Experimental Implementation}

\begin{figure}
\includegraphics[width=1\columnwidth]{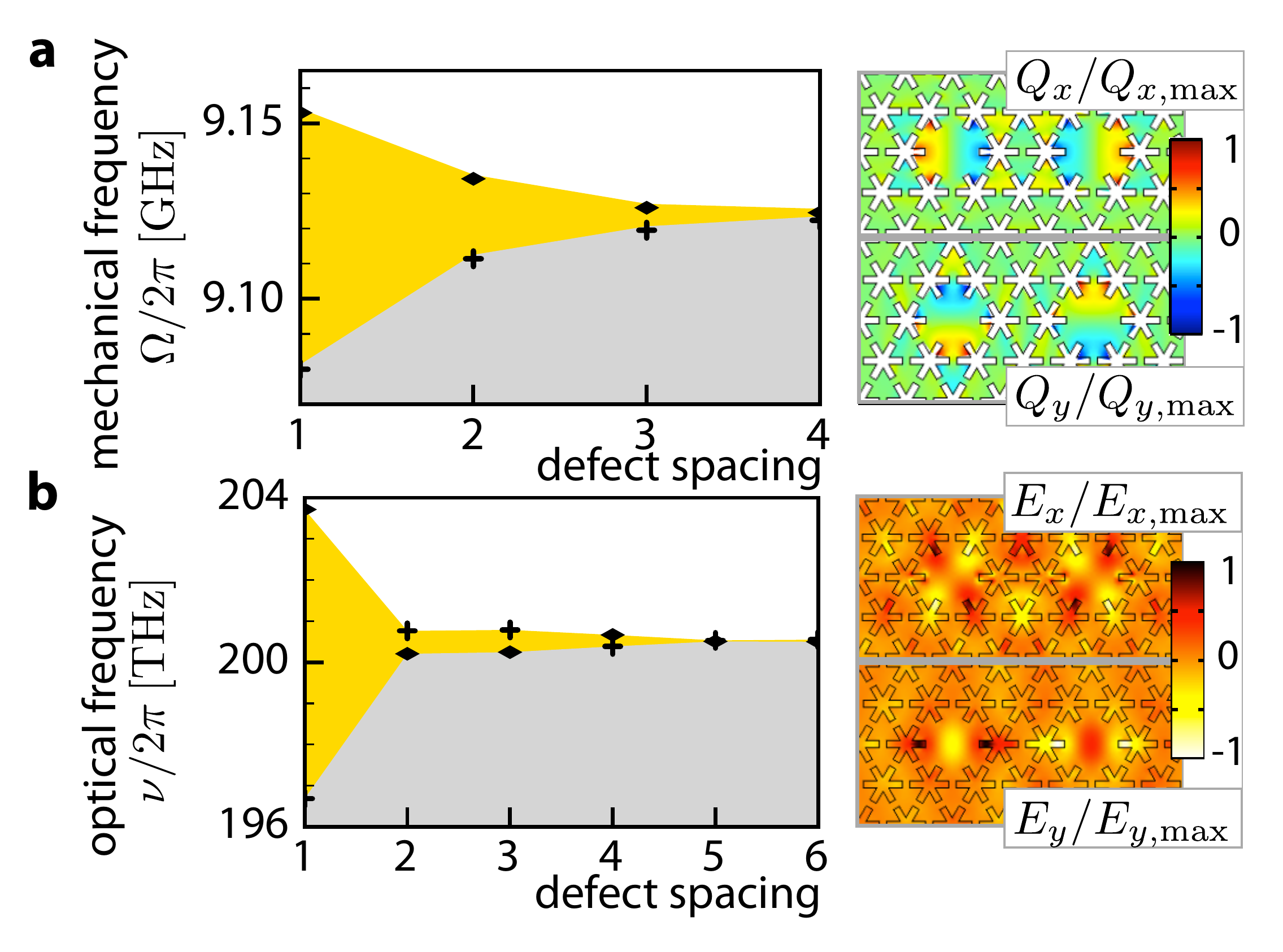}\caption{Variation of coupling strength with separation between defect cavities:
(a) mechanical and (b) optical frequencies of symmetric (cross) and
antisymmetric (diamond) normal modes for two defects on a snowflake
optomechanical crystal (as discussed in the main text). The insets
show the displacement field (electrical field) components of the antisymmetric
mode for a defect separation of $2$. The bare frequencies of the
localized eigenmodes are $9.12\,{\rm GHz}$ and $200.5\,{\rm THz}$,
respectively. We note that the optical splittings (and likewise $J$)
change sign when varying the defect separation. Results obtained from
finite element simulations.}
\label{FigS3}
\end{figure}

(a) \textit{Competing setups - }Here, we provide some details on possible
experimental implementations suitable for observing the mechanical
transition:
\begin{itemize}
\item Microdisks \cite{2006_Barclay_IntegrationOfFiberCoupledMicrodisks,2010_Ding_HighFrequencyResonator,2012_Zhang_SynchronizationOptomech}
and microtoroids \cite{2003_Armani_ToroidMicrocavityOnAChip,2007_Armani_MicrocavityArrays,2012_Verhagen_QuantumCoherentCoupling}
fabricated on microchips: Strong optical coupling between resonators
is feasible via evanescent fields \cite{2012_Zhang_SynchronizationOptomech}.
For both types of setups, scalability still has to be shown.
\item Micro- and nanomechanical beams \cite{2008_Regal_MicrowaveCavity,2010_Rocheleau_PreparationAndDetection}
or membranes \cite{2011_Teufel_SidebandCooling_Nature} coupling to
superconducting microwave cavities: Mechanical interaction may be
achieved via a common support or, capacitively, by applying a voltage
bias between the mechanical resonators. Two-dimensional arrays of
coupled microwave cavities are starting to be developed \cite{2012_Houck_On-ChipSimulation}.
Related electromechanical systems (see, e.g., \cite{2006_Naik_CoolingNanomechResonator}
for a setup comprising a nanobeam coupling to a superconducting single
electron transistor) may also be employed.
\item Optomechanical crystals \cite{2011_Chan_LaserCoolingNanomechOscillator,2010_Safavi_2DSimultaneousBandgap,2011_Gavartin_2DOptomechanicalCrystal}
feature small mode volumes and are thus very suitable for extensions
to optomechanical arrays. Some details of the proposed implementation
are given below (Fig. \ref{FigS3}). 
\end{itemize}
(b) \textit{Required parameters: }According to our semiclassical analysis,
the essential requirement is 
\begin{equation}
K\bar{A}^{2}\gtrsim\Gamma_{{\rm opt}}>\Gamma n_{{\rm th}}.\label{eq:Req_params}
\end{equation}
In this case, quantum noise ($\Gamma_{{\rm opt}}$) dominates over
thermal fluctuations ($\Gamma n_{{\rm th}}$), which enter the model
by replacing $\Gamma\to\Gamma n_{{\rm th}}\approx k_{B}T/Q$ in Eqs.
(\ref{eq:EoM}). The mechanical transition can then be studied by
varying $\Gamma_{{\rm opt}}$ via the laser detuning $\Delta$.

Recent experiments have demonstrated $\Gamma_{{\rm opt}}>\Gamma n_{{\rm th}}$,
see Table \ref{Table1}. We note that two-dimensional optomechanical
crystal devices are expected to show very good optomechanical properties
\cite{2010_Safavi_2DSimultaneousBandgap}, even exceeding those of
existing one-dimensional setups \cite{2011_Chan_LaserCoolingNanomechOscillator}.

(c) \emph{Photon and phonon hopping amplitudes}: Using finite element
simulations, we studied the hybridization of the photon and phonon
modes of two defect cavities inside a two-dimensional snowflake silicon
optomechanical crystal to obtain the hopping constants as a function
of defect separation. Parameters: relative permittivity $\epsilon_{r}=11.68$,
Poisson ratio $0.17$, density $2329\,{\rm kg/m^{3}}$, Young's modulus
$170\,{\rm GPa}$, snowflake design \cite{2011_Safavi-Naeini_PPT}
with lattice constant $500\,{\rm nm}$, snowflake radius $168\,{\rm nm}$
and snowflake width $60\,{\rm nm}$. The mechanical and optical splittings
(i.e. couplings $2K/z$ and $2J/z$, respectively) reach values of
up to $8\,\%$ and $4\,\%$ of the mechanical and optical eigenfrequencies,
respectively (Fig. \ref{FigS3}). These values are compatible with
the requirements given by Eq. (\ref{eq:Req_params}), even for very
small oscillation amplitudes of the order of the mechanical zero-point
width, $\bar{A}\approx1$.

Other experimental approaches may realize different coupling terms,
 e.g. $(\hat{x}_{i}-\hat{x}_{j})^{2}$ \cite{2002_Buks_MicromechArray}.
Additional fast rotating terms like $\hat{b}_{i}\hat{b}_{j}$ are,
however, negligible for $K/z\ll\Omega$.

We note that the transition from incoherent to synchronized dynamics
is also observable for extended optical modes ($J\gg\Omega,\kappa$),
see Fig. \ref{FigS1}(c).

(d) \textit{Optical drive: }In principle, the methods used for driving
single defect modes via tapered and dimpled fibers \cite{2007_Michael_OpticalFiber-TaperProbe}
and optical waveguides \cite{2013_Safavi-Naeini_SqueezingOfLight,2013_Cohen_OpticalCoupling}
can be extended to larger scales. To limit experimental efforts, however,
one may drive the array using a freestanding broad laser beam, see
the schematic picture in Fig. \ref{Fig1}. The coupling between laser
mode and the in-plane cavity modes will be relatively weak, which,
however, can be compensated via the laser power. Preliminary experimental
results on free space coupling to optomechanical crystals have been
reported in \cite{2012_Chan_PhDThesis}.

(e) \textit{Detection: }To detect the mechanical transition, measurements
of the optical field emitted from the array are sufficient, see Fig. \ref{Fig3}(d).
To determine correlations between two separate lattice sites, the
intensity emanating from these defect cavities may be probed by two
tapered fibers in the near-field of the selected cells \cite{2007_Michael_OpticalFiber-TaperProbe}
or by especially designed waveguides \cite{2013_Safavi-Naeini_SqueezingOfLight,2013_Cohen_OpticalCoupling}.
The correlations between the mechanical sideband components of the
optical intensities $\mathcal{I}_{i}(t)=\int_{t}^{t+2\pi/\Omega_{{\rm eff}}}e^{i\Omega_{{\rm eff}}t'}|\alpha_{i}|^{2}(t')dt'$,
are then proportional to the mechanical correlations, 
\begin{equation}
\langle\mathcal{I}_{i}\mathcal{I}_{j}^{\star}\rangle\propto\langle e^{i\varphi_{i}}e^{-i\varphi_{j}}\rangle.\label{eq:Corr_optical}
\end{equation}
\begin{figure}[b]
\includegraphics[width=1\columnwidth]{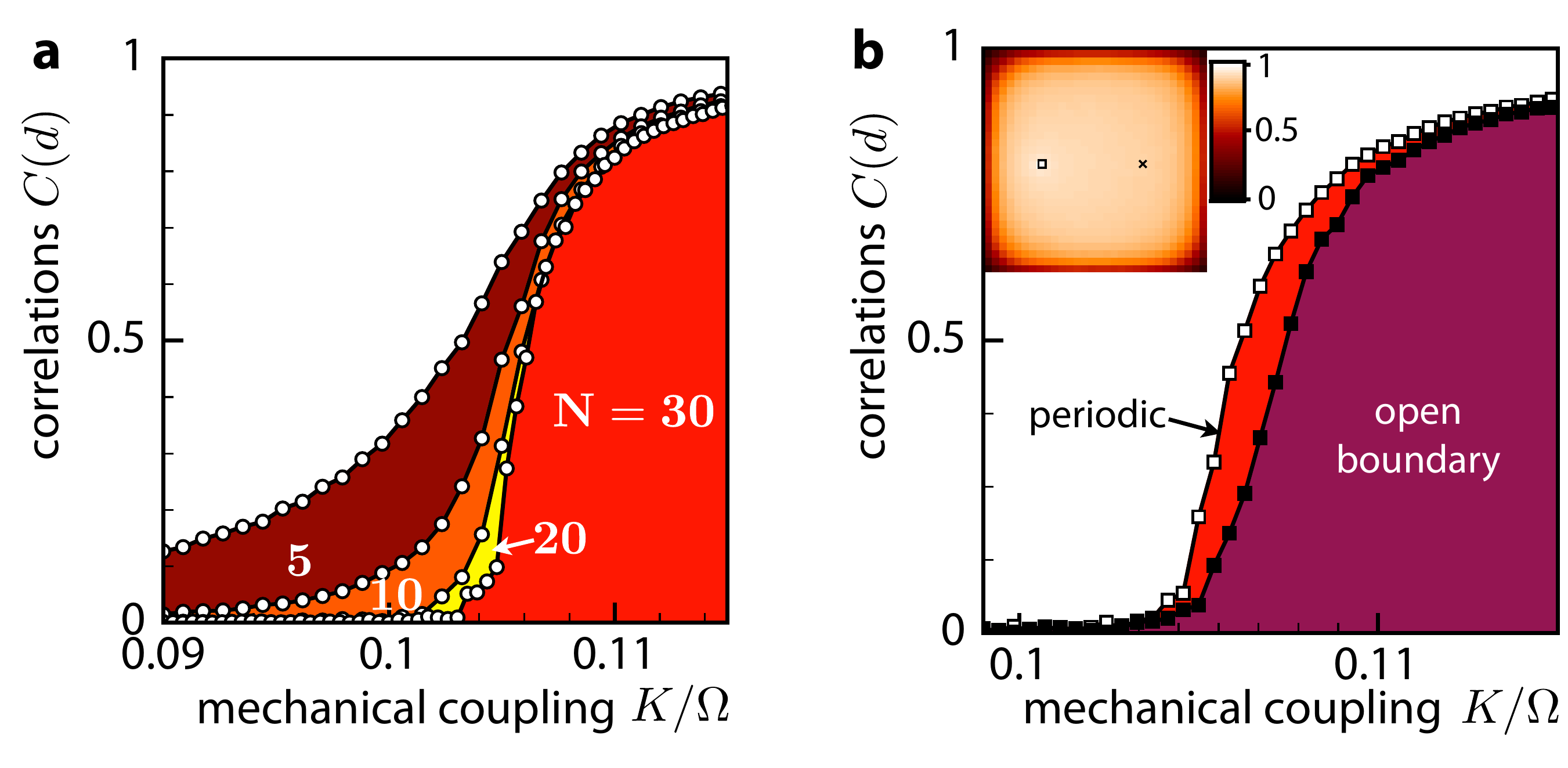}

\caption{Langevin dynamics on finite square lattices, confirming results shown
in Fig. \ref{Fig4}: (a) Correlations $C(d)$ as a function of coupling
strength for different lattice sizes $N\times N$ with $N=\{5,10,20,30\}$
($d=\{2,4,9,14\}$, periodic boundary conditions). (b) Correlations
$C(d=14)$ on a $30\times30$ square lattice with periodic boundary
conditions (empty squares) and open boundary conditions (filled squares).
The inset shows the correlations $|\langle e^{i\varphi_{i}}e^{-i\varphi_{j}}\rangle|$
for $i=(8,15)$ (square) and $K=0.115\,\Omega$ and open boundary
conditions. The cross marks the lattice site $j=(22,15)$ for which
the correlations are shown in the main plot. Other parameters as in
Fig. \ref{Fig4}.}

\label{FigS2}
\end{figure}

\section{Numerical Methods}

To study the quantum dynamics of the system, we numerically integrate
the Lindblad master equation (see main text) to the steady state (independent
of initial conditions) using a fourth-order Runge-Kutta algorithm.
The size of the Hilbert space is optimized to enable an adequate representation
of the physical state while obtaining reasonable simulation times.
Typically, $10-20$ photon and phonon levels, respectively, were taken
into account. When slowly sweeping through parameter space, bistable
behavior is revealed. This can be seen most prominently in Fig. \ref{FigS1}(c)
from the cut line in the right part of the plot.

The numerical integration of the Langevin equations (\ref{eq:EoM})
was obtained using a fourth order Runge-Kutta method for stochastic
differential equations \cite{1995_Kasdin_DiscreteSimulation_SDE}.
When evaluating the dependence of the threshold value on the quantum
parameter $g_{0}/\kappa$ (Fig. \ref{Fig4}), the value
of $g_{0}\alpha_{L}$ was held constant. In this case, only the strength
of quantum fluctuations changes (keeping the classcial solution constant).
For very large values of $g_{0}/\kappa$, fluctuations are overestimated
by the Langevin equations and one has to rely on the exact quantum
simulations.

The finite element simulations of Figs.\ref{Fig1} and \ref{FigS3}
were performed using COMSOL Multiphysics. We studied a hexagonal silicon
slab with sides of length $5.6\,\mu m$ and restricted our simulations
to two space dimensions, i.e. in-plane elastic deformations and electromagnetic
waves, see also \cite{2011_Safavi-Naeini_PPT}. Extended simulations
in three dimensions have to be employed to analyze the coupling of
the cavity modes to the out-of plane-modes.

\end{document}